\documentclass[aps,prl,reprint,floatfix,amsmath,amssymb,twocolumn]{revtex4-2}
\usepackage{CJK}
\usepackage{color}
\usepackage{epsfig}
\usepackage{graphicx}
\usepackage{dcolumn}
\usepackage{bm}
\usepackage[colorlinks=true,citecolor=blue, dvipdfm]{hyperref}
\usepackage{amssymb}
\usepackage{enumerate}

\begin{document}
\title{{When irrelevance governs: dissipation-regression scaling in open quantum criticality}}

\author{Fan Zhong}
\email{stszf@mail.sysu.edu.cn}
\affiliation{School of Physics and State Key Laboratory of Optoelectronic Materials and
Technologies, Sun Yat-sen University, Guangzhou 510275, People's Republic of China}

\date{\today}

\begin{abstract}
{The real-time dynamics of quantum phase transitions in open systems coupled to finite-temperature environments sits at the heart of contemporary condensed-matter and quantum-information science. Here, invoking a Schwinger–Keldysh formalism, we uncover that, within the coherence-dominated regime of critical real-time evolution, a dissipative coupling deemed naively irrelevant determines, counterintuitively, the longest timescale that holds sway. Possessing a well-defined temporal scaling dimension, it engenders a novel dissipation-regression scaling that systematically reproduces the unusual Kibble–Zurek scaling and relaxation observed in recent numerical studies. By contrast, dissipation-dominated dynamics behaves qualitatively differently. There, furthermore, we show that the discretized and diagonalized formulations of the same open quantum Ising chain realize distinct universality classes.}
\end{abstract}


\maketitle

Quantum phase transitions (QPTs) describe fundamental changes in the ground state of a many-body system driven by quantum fluctuations at absolute zero~\cite{sachdev}. Unlike classical transitions, static and dynamic behaviors are inherently intertwined in QPTs~\cite{Hertz}. Real-time dynamics of QPTs are therefore central to modern condensed matter physics, statistical physics, and quantum computing, among others. While manipulating isolated real-time dynamics of QPTs is technically possible in some deliberately designed setups~\cite{Greiner,Kinoshita,Hofferberth}, real experiments generally inevitably involve coupling to an environment. Moreover, the absolute zero limit is thermodynamically inaccessible, making the role of finite temperature unavoidable. Therefore, probing, understanding, predicting, and controlling QPTs in open systems is of pivotal importance.

Continuous QPTs constitute one prominent class of QPTs. The associated quantum criticality shares scaling and universality as hallmarks with its classical counterpart~\cite{Mask,Cardyb,Justin,Amit}. An effective method to study such collective behavior is to drive a system through its critical point by varying the temperature or a parameter in the system Hamiltonian in time. The renormalization-group (RG) theory for linear driving was originally developed for first-order phase transitions~\cite{zhongl05}, building on the tradition of applying the method to study dynamics of dislocations (via K\^{e}'s pendulum~\cite{Kep}) and first-order phase transitions (see, e.g., Refs.~\cite{Ke,Zhang,Zhang1,Zhang3,zhongprl}). The theory was later extended to classical critical phenomena~\cite{zhongxu,zhong06,Gong}. This results in a theory of finite-time scaling (FTS)~\cite{Gong,Gong1,Huang} since the rate of the driving imposes a controllable finite timescale on the evolution of the system, in close analogy to the role that the finite system size plays in the well-known finite-size scaling~\cite{fss,Brezin,Brezin1,Amit,Cardyb,Justin}. FTS enables one to describe the scaling of the entire driving process around the critical point and to detect both equilibrium and nonequilibrium critical properties. An RG theory for a nonlinear driving with multiple parameters of arbitrary form has also been developed, leading to a series of driven nonequilibrium critical phenomena such as hysteresis, negative susceptibility, and crossover between distinct regimes dominated by the different parameters in the driving~\cite{Feng}.

A driven nonequilibrium critical phenomenon is the generation of topological defects via the famous Kibble-Zurek mechanism. Originally proposed in cosmology~\cite{Kibble1,Kibble2} and later applied to condensed-matter physics~\cite{Zurek1,Zurek2}, the mechanism assumes a frozen-in diabatic regime sandwiched between two adiabatic regimes during cooling through a critical point. The resulting topological defect density is estimated from the boundary between the regimes to obey a Kibble-Zurek scaling, a power law of the cooling rate with an exponent related to both equilibrium and dynamic critical exponents. Despite extensive study~\cite{revqkz1,revqkz2,Zurek4}, several important questions remain unanswered: what the accuracy of the scaling is, how other parameters (e.g., an external field) influence it, and even whether the specific temperatures at which the defect density is estimated affect it. For instance, although the scaling is  exactly derived at the boundary, which varies with the cooling rate, the defects are sometimes directly estimated at a fixed temperature regardless of the cooling rate~\cite{Laguna1,Braun,Suzuki}. Moreover, although the topological defects themselves may be important, they are not good observables, at least in some cases. An example is the ferromagnetic Ising model where the interface between up and down spins is a topological defect. At finite temperature, the vigorous fluctuation of spins renders any meaningful estimation of topological defects unreliable. However, all these questions and others, such as how observables other than the defect density  (e.g., an order parameter) scale, are transparent in FTS, as will be seen below, together with a case in which the defect density may be an inappropriate starting point in perturbation expansions.

Returning to quantum criticality in open quantum systems, 
a strong enough coupling to the bath can trigger a new dissipative phase transition~\cite{Sudip,Weiss,Wang}, while new classes of quantum criticality different from the nondissipative ones emerge independent of the coupling strength~\cite{Werner,Sper,Sper12,Oshiyama}. The first investigation of a system driven through its own critical point at a fixed finite bath temperature yielded, via heuristic and scaling arguments, different expressions for the excitation density of coherent and incoherent contributions and a crossover between them~\cite{patane,patane1}. A systematic theoretical framework, based on the Lindblad equation~\cite{Lindblad,attal,Breuer}, was later proposed and tested using FTS~\cite{Yin,Yin3}. The key feature of the theory is the introduction of the coupling as a new indispensable variable, in addition to the temperature and the original scaling variables of the closed system, aiming to probe quantum criticality at finite temperature. The {same} framework was then further validated using finite-size scaling~\cite{Yinchen,Rossini}. Recently, an unusual Kibble-Zurek scaling of the excitation density~\cite{Nalbach,Kuo,Bacsi,King,Ding,Kriel} that differs from the standard one~\cite{Zurek,Polk,{Barankov,Sen}} has been obtained at varying temperatures, across various models, driving protocols, and bath spectral densities, without resorting to finite-size scaling, partly due to technical advances in solving master equations~\cite{Prosen}. 

Here, {focusing on quantum critical} systems coupled to finite-temperature environments {featuring a power-law spectral density with an exponent $s$, we first show that neither perturbative expansions nor crossover arguments suffice to account for the unusual Kibble–Zurek scaling observed within the existing framework---underscoring the need for a new conceptual ingredient. Then, invoking a nonequilibrium Schwinger-Keldysh field theory, we demonstrate that within the coherent regime---where unitary dynamics would otherwise dominate---the naively irrelevant dissipative coupling determines, counterintuitively, the longest timescale, thereby dictating the asymptotic evolution. It resets the timescale with its own distinct RG eigenvalue, yielding what we term dissipation-regression scaling (DRS). Crucially, DRS is not an instance of the familiar dangerously irrelevant variables---which typically result in violation of hyperscaling above the upper critical dimension~\cite{Fisher82,AmitPeliti,Zenged} or generation of emergent symmetries in the ordered phase~\cite{Nelson,Oshikawa,Lou,Okubo,Leonard,Ding16,Nakayama,Yang,Yang1,Chen,Shu,Shu26}---but instead operates purely within the dynamical sector. As such, DRS multiplicatively rescales the dynamic critical exponent, thus providing a unified mechanism behind the unusual Kibble–Zurek scaling and relaxation behavior reported. Furthermore, we unexpectedly find that two common representations of the open quantum Ising chain---its diagonalized and discretized forms---belong to mutually inequivalent universality classes in the dissipative regime, a discrepancy highlighting the subtlety of defining real-time critical dynamics in open quantum systems.}

{To set the stage, we first} consider a system with Hamiltonian ${\mathcal H}$ coupled to a bath, described by a Lindblad equation:~\cite{Lindblad,attal,Breuer},
\begin{equation}
	\frac{\partial \rho}{\partial t}=-i[{\cal H},\rho]+\lambda\left( [{\cal L}\rho, {\cal L}^{\dagger}]+[{\cal L}, \rho{\cal L}^{\dagger}]\right), \label{drho}
\end{equation}
where $\rho$, $t$, and $\lambda$ are the system density matrix, time, and coupling constant, respectively, ${\cal L}$ denotes a set of jump operators {that exchange quasiparticles with the bath}, the superscript ${\dagger}$ is the Hermitian conjugate, and square brackets represent commutators. Planck constant $\hbar$ and Boltzmann constant $k_{\rm B}$ are set to unity. The bath is initially thermalized at temperature $T$ and is characterized by a spectral density $J(\omega)$ {$\propto$ $\omega^s\Theta(\omega_c-\omega)$ up to a proportionality constant absorbed into the coupling $\lambda$, where} the cutoff frequency {$\omega_c$ lies} well above the mode frequencies $\omega$ {and $\Theta(x)$ is the Heaviside step function}. The exponent $s$ {$(\geq0)$} specifies the nature of the dissipative bath: $s=1$ represents an Ohmic bath, $0<s<1$ indicates a sub-Ohmic bath, while $s>1$ corresponds to a super-Ohmic bath~\cite{Caldeira,Leggett}. {As numerical results both from Eq.~(\ref{drho}) for different models~\cite{Bacsi,Ding,King,Kriel} and from the analogue equation~\cite{Kuo} are already available, we need not re-calculate them here. Accordingly, we do not specify the explicit form of ${\cal L}$, noting only its proportionality to 
$\sqrt{J(\omega)}$, with the necessary summations understood.}

{Equation~(\ref{drho}) was first derived microscopically for a quantum transverse-field Ising model linearly coupled with a set of independent bosonic heat bath at $s=0$, under the weak-coupling, Born-Markov, and secular approximations~\cite{Yin3}. It can be rigorously constructed for open quadratic quantum systems of either fermionic or bosonic models with a general $J$ under the same approximations~\cite{dabbru}. Although non-Ohmic baths generally give rise to non-Markovian memory effects, Eq.~(\ref{drho}) is often phenomenologically employed to study properties of open quantum systems~\cite{Bacsi,King,Ding,Kriel,Sieberer}. Note, however, that the theory to be developed does not rely on this equation; rather, we partly utilize the results to verify the theory.}

We next recapitulate the essence of the scaling hypothesis. For a single critical point at $g=0$, the scaling hypothesis for the excitation density ${\mathfrak D}$ in a $d$-dimensional space is~\cite{Yin,Yin3}
\begin{equation}
	{\mathfrak D}(t,g,T,\lambda,L)= b^{-d}{\mathfrak D}(tb^{-z}, gb^{1/\nu}, Tb^{z},\lambda b^{(1-s)z}, L^{-1}b ),\label{sgtb}
\end{equation}
where $b$ is a length rescaling factor, $L$ the system size, and $\nu$ and $z$ are the critical exponents for the correlation length and the dynamics, respectively~\cite{Mask,Cardyb,Justin,Amit}. We have included the size of the system $L$ to incorporate finite-size effects and have neglected dimensional factors for simplicity. Equation~(\ref{sgtb}) can be derived from the RG theory in classical critical phenomena when the scaling form includes the variables $t$, $g$, $L$, and other quantities such as an order field. In the quantum context, the same equation can also be derived in the presence of $g$ and $T$, because $1/T$ plays the role of imaginary time~\cite{Hertz,sachdev}. With $\hbar=k_{\rm B}=1$, frequency and energy share the same RG exponent as $T$. This determines the scaling dimension or RG eigenvalue for $\lambda$ in Eq.~(\ref{sgtb}), since $\lambda$ itself is the inverse time. {Application of Eq.~(\ref{sgtb}) to} $s=0$ has been studied previously~\cite{Yin,Yin3,Yinchen,Rossini}.  

{To expose the physics encapsulated in} the scaling hypothesis Eq.~(\ref{sgtb}){, we sets $b=|g|^{-\nu}$, resulting in}
\begin{equation}
    {\mathfrak D}= |g|^{d\nu}f(t|g|^{\nu z}, T|g|^{-\nu z},\lambda |g|^{-(1-s)\nu z}, L^{-1}|g|^{-\nu} ),\label{sgtxi}
\end{equation}
which {includes corrections such as time and size and is thus} a quasi-equilibrium scaling form, where $f(x_1,x_2,x_3,x_4)={\mathfrak D}(x_1,1,x_2,x_3,x_4)$ is a universal scaling function. The last argument of $f$ indicates that $|g|^{-\nu}$ is a length scale. Indeed, it is simply the correlation length $\xi\sim|g|^{-\nu}$. Accordingly, for a given $g$ and a certain value of ${\mathfrak D}$, the first argument of $f$ determines the corresponding correlation time $\zeta_{\rm corr}\sim|g|^{-\nu z}\sim\xi^z$ and its defining dynamic critical exponent {$z$}.
Moreover, the leading behavior of Eq.~(\ref{sgtxi}) becomes asymptotically ${\mathfrak D}\sim\xi^{-d}$, which is {related to} the conventional definition of the topological defect density. This asymptotic behavior is reached when all scaled variables are assumed to be vanishingly small. For example, $L^{-1}|g|^{-\nu}\ll1$, i.e., $\xi\ll L$, correctly quantifies the thermodynamic limit. In addition, $\lambda|g|^{-(1-s)\nu z}\ll1$ and $T|g|^{-\nu z}\ll1$ are equivalent to $\lambda\ll\xi^{-z}\sim\zeta_{\rm corr}^{-1}$ (for $s=0$) and $T\ll\Delta$, respectively, where $\Delta\sim|g|^{\nu z}\sim\xi^{-z}$ is the energy gap above the ground state. These two inequalities indicate that $\lambda$ and $T$, which are associated with dissipation and thermal excitation, become negligible {when they are much smaller than the inverse correlation time and the energy gap, respectively. Note that they may dominate sufficiently close to the quantum critical point because they are both relevant parameters in the RG sense.}

Now we consider real-time driving across the quantum critical point by nonlinearly ramping a control parameter, for example,
\begin{equation}
g=Rt^n,\label{grt}
\end{equation}
where $n$ is a constant exponent and $R>0$ is the rate of the ramp. Let $r$ be the RG eigenvalue for $R$, implying that at scale $b$, $R$ becomes ${R'=}~Rb^{r}$. The RG theory predicts that~\cite{Feng}
\begin{equation}
	r=nz+1/\nu.\label{rznu}
\end{equation}  
Since only two out of the trio $g$, $R$, and $t$ are independent, we replace $t$ by $R$ in Eq.~(\ref{sgtb}) and set $b=R^{-1/r}$, yielding an FTS scaling form~\cite{Gong,Gong1,Huang},
\begin{equation}
	{\mathfrak D}= R^{d/r}f_R(gR^{-1/r\nu}, TR^{-z/r},\lambda R^{-(1-s)z/r}, L^{-1}R^{-1/r} ),\label{sgtr}
\end{equation} 
with finite-size corrections, where $f_R$ is another scaling function. Throughout, exponents such as $1/r\nu$ are understood as $1/(r\nu)$ to reduce parentheses. The leading term $R^{d/r}$ in Eq.~(\ref{sgtr}) is simply the standard Kibble-Zurek scaling, which can only be accurately achieved when all the scaled variables in $f_R$ are either constant or negligible. In particular, the condition $\hat{g}R^{-1/r{\nu}}=1$ precisely defines the frozen value $\hat{g}$~\cite{Huang}. {Consequently}, if $g$ is constant (as opposed to $\hat{g}$, which varies with $R$ {as $R^{1/r\nu}$}), then $f_R$ still depends on $R$. As a result, ${\mathfrak D}$ is not {strictly} proportional to $R^{d/r}$, even if all remaining variables {are} neglected. {Therefore, it is misleading to claim that the Kibble-Zurek scaling depends solely on the equilibrium critical exponents.} In addition, it is clear that $R^{-z/r}$ is a controllable driving timescale, which is why Eq.~(\ref{sgtr}) is called the FTS form.

In the standard Kibble-Zurek scaling derived from the leading behavior in Eq.~(\ref{sgtr}), the dissipation term involving $\lambda$ acts only as a perturbation, i.e., $\lambda R^{-(1-s)z/r}\ll1$, or equivalently, $\lambda^{-1}\gg R^{-(1-s)z/r}$. This implies that the dissipation time is much longer than the driving time and can therefore be neglected. To find a Kibble-Zurek scaling that is {related} to $\lambda$, we can expand the scaling function $f_R$ with respect to this {small} variable, resulting in
\begin{equation}
	{\mathfrak D}(R,g,T,\lambda,L)={\mathfrak D}_{\rm coh}+{\mathfrak D}_{\rm inc},\label{sgtkz}
\end{equation}
where the coherent contribution to the excitation density is
\begin{eqnarray}
   {\mathfrak D}_{\rm coh}\sim R^{\sigma_{\rm coh}},\qquad\quad\label{ncoh}\\
   \sigma_{\rm coh}=d/r=d\nu/(1+n\nu z),\label{scoh} 
\end{eqnarray}
and the incoherent one is ${\mathfrak D}_{\rm inc}\sim \lambda R^{\sigma_{\rm inc}}$ with the Kibble-Zurek scaling exponent $\sigma_{\rm inc}=[d-(1-s)z]/r=[d-(1-s)z]\nu/(1+n\nu z)$. Equation~(\ref{sgtkz}) explicitly justifies the separation of the two contributions to ${\mathfrak D}$, while Eq.~(\ref{scoh}) for $\sigma_{\rm coh}$ is the standard Kibble-Zurek scaling {under linear~\cite{Zurek,Polk} and nonlinear driving~\cite{Barankov,Sen}}.
For the $d=1$ QPTs in the transverse Ising model~\cite{Pfeuty} and in the critical line of the Kitaev model~\cite{Kitaev}, where $\nu=z=1$, we find $\sigma_{\rm inc}=sz/(1+n)$, different from the standard coherent exponent $\sigma_{\rm coh}=1/(1+n)$. Yet, this $\sigma_{\rm inc}$ has no physical counterpart{---a striking instance being $\sigma_{\rm inc}=0$ at $s=0$ for the aforementioned models---} except for $s=1$, where $\lambda$ is dimensionless (see below).

To obtain a better estimate of $\sigma_{\rm inc}$, we can expand the correlation length $\xi$ and employ ${\mathfrak D}\sim\xi^{-d}$. The FTS form for $\xi$~\cite{Zuo} is analogue to that for ${\mathfrak D}$ given in Eq.~(\ref{sgtr}); the only modification is the replacement of the leading asymptotic behavior by $R^{-1/r}$, a driving length scale. Therefore
\begin{eqnarray}
	{\mathfrak D}_{\rm inc}\sim \lambda^{-d}R^{\sigma_{\rm inc}},\qquad\qquad\quad\label{ninc}\\
	\sigma_{\rm inc}=\frac{d[1+(1-s)z]}{r}=\frac{d\nu[1+(1-s)z]}{(1+n\nu z)}.\label{sinc} 
\end{eqnarray}
{For the two specified models, $\sigma_{\rm inc}=(2-s)/(1+n)$. This result is consistent with numerical data both at $s=0$ (with $n\nu=1$) and at the special case of $s=1$ mentioned above~\cite{Bacsi} [cf. Eq.~(\ref{sg}) below]. The remarkable $\lambda^{-d}$ factor for any $s$ in Eq.~(\ref{ninc}) likewise matches the numerical findings~\cite{Bacsi}, though it} differs from the linear factor in the direct expansion of Eq.~(\ref{sgtr}). It is {precisely} this factor that invalidates the direct expansion of ${\mathfrak D}${, unlike that of $\xi$}. {Despite its apparent divergence as $\lambda\to0$, this factor is consistently obtained in all related studies~\cite{Bacsi,King,Ding,Kriel,Kuo}. Within the present framework, only the special case $s=1$ permits a physical explanation, to which we will return in the theory developed below. This reveals another limitation of the present framework.} Note that the expressions for $\sigma_{\rm inc}$ and $\sigma_{\rm coh}$ differ, as seen from Eqs.~(\ref{scoh}) and~(\ref{sinc}). Their crossover occurs at $\lambda R^{-z/r}\approx 1$ for $s=0$, when the two timescales involved, the dissipation time $\lambda^{-1}$ and the driving time $R^{-z/r}$, become comparable. The above result illustrates that the density of the topological defects may be an unsuitable quantity to start with.

{We have exploited the direct Taylor expansion in the correlation length $\xi$ to obtain the excitation density for weak coupling above. However, for strong coupling, namely, when $\lambda R^{-(1-s)z/r}\gg 1$, a crossover to the coupling-dominated regime occurs as in a conventional scaling analysis~\cite{Huang,Zuo}. This can be obtained again by replacing $t$ by $R$ in an equation similar to Eq.~(\ref{sgtb}) for $\xi$ and setting now $b=\lambda^{-1/(1-s)z}$, yielding a scaling form,
\begin{equation}
	\begin{split}
	{\xi}= \lambda^{-1/(1-s)z}f_{\lambda}(g\lambda^{-1/(1-s)\nu z}, T\lambda^{-1/(1-s)},~\quad\\
	 R\lambda^{-r/(1-s)z}, L^{-1}\lambda^{-1/(1-s)z} )\label{sgtla}
    \end{split}
\end{equation} 
with $f_{\lambda}$ being a new scaling function. Now, we can again expand $f_{\lambda}$ in the small $R\lambda^{-r/(1-s)z}$, leading to ${\mathfrak D}_{\rm inc}\sim \lambda^{(1+r)d/(1-s)z}R^{-d}$, which contradicts the correct results, Eqs.~(\ref{sgtbrr}) and~(\ref{sgrr}) below. One can convince oneself that the correct results can never be obtained from direct consideration of the Taylor expansion of ${\mathfrak D}$ itself, or of the finite-size or finite-temperature effects within the framework.}

{We have shown that perturbation theory yields the expected Kibble–Zurek exponents only in two isolated cases: $s=0$ (with $n\nu=1$) and $s=1$. Moreover, even these two successful cases suffer from a divergence as $\lambda\to0$, indicating a breakdown of the perturbative expansion itself.} As can be seen from Eq.~(\ref{sgtb}), at $s=0$, $\lambda$ {shares the scaling dimension of} $t^{-1}$, whereas at $s=1$, it is dimensionless and serves purely as a time unit. Conversely, the failure of the standard timescale $b^{-z}$ for $s\neq 0, 1$ indicates that this choice of the scale is inadequate. {We now introduce a conceptually different element absent from the preceding discussion: DRS, which supplies the missing characteristic timescale.} 

{To this end, we turn to a general nonequilibrium quantum field-theoretic framework~\cite{Kamenev}. For definiteness, rather than continuing with the two aforementioned models, we adopt the canonical bosonic quantum field theory given by
\begin{eqnarray}
	{\cal H}=\int d\mathbf{x}\left[\frac{1}{2}\Pi^2+\frac{1}{2}(\nabla\phi)^2+ \frac{g}{2}\phi^2+\frac{u}{4!}\phi^4\right],\nonumber\\
	\left[\phi(\mathbf{x}),\Pi(\mathbf{x}')\right]=i\delta(\mathbf{x}-\mathbf{x}'),\label{qft}\qquad\quad
\end{eqnarray}
since they fall into the same unitary universality class, where $\Pi$ denotes the momentum conjugate to the field $\phi$ and $u$ measures the self-interaction strength. To study the real-time evolution, following the same strategy that led us from the Liouville equation to Eq.~(\ref{drho}), we linearly couple the system to a collection of independent bosonic heat baths in the Caldeira-Leggett form~\cite{Caldeira,Leggett}. However, in contrast to relying on the Markovian approximation, we employ the Schwinger-Keldysh technique~\cite{Sieberer,Kamenev,Breuer,Keldysh} to derive the Keldysh effective action as 
\begin{eqnarray}
			S[\phi_c,\phi_q]
			= \int_k \Big[ \phi_c^*\mathcal{G}_R^{-1}\,\phi_q
			+ \phi_q^*\mathcal{G}_A^{-1}\phi_c 
			- \frac{i}{2}\Sigma_K \phi_q^*\phi_q\Big]\quad\nonumber\\
			- \frac{u}{12}\int_{\{k\}}\Big[
			\phi_c^3\phi_q+ \phi_c\phi_q^3
			\Big],\qquad\qquad~\label{sphi}\\
\mathcal{G}_R^{-1} = \omega^2-\mathbf{k}^2-g
- i\lambda\omega^s\Theta(\omega_c-\omega),\qquad\qquad\nonumber\\
\mathcal{G}_A^{-1} =\left(\mathcal{G}_R^{-1}\right)^*\!,~
\Sigma_K=-i\left(\mathcal{G}_R^{-1}-\mathcal{G}_A^{-1}\right)\coth\!\left(\frac{|\omega|}{2T}\right)\!,~\quad\label{gram1}
\end{eqnarray}
where $\phi_c$ and $\phi_q$ are the Keldysh classical and quantum fields, respectively, asterisks mark complex conjugates, the integration over $k$ means $\int d\omega d\mathbf{k}/(2\pi)^{d+1}$, and $\{k\}$ denotes a set of four $k$ constrained by total energy-momentum conservation. In Eq.~(\ref{gram1}), we have absorbed the dissipation-induced shift of the critical point in $g$ for clarity.}

{From Eq.~(\ref{gram1}), it is evident that independent of $\lambda$, $s=2$ separates two different regimes. For $s>2$, $\omega^2\gg\omega^s$, the low-energy behavior is governed by $\omega^2$. One can directly read off $z=1$ for both the real time and imaginary time $1/T$ from the dispersion-related relation $\omega^2-\mathbf{k}^2$ and $\omega/2T$, respectively, in the equation. Consequently, coherent dynamics dominates and criticality is described by Eq.~(\ref{sgtb}), with the scaling dimension of $\lambda$ acquiring a modified value $(2-s)z$ due to $\omega^2$; hence $\lambda$ stays irrelevant for $s>2$. This is the conventional unitary quantum Ising universality class.}
	
{Counter-intuitively, however, the imaginary part of $\mathcal{G}_R^{-1}$ yields a quasi-particle lifetime of order $\lambda^{-1}|\mathbf{k}|^{-sz}$, which, for $s>2$ and $z=1$, is much longer than the coherent time $|\mathbf{k}|^{-z}$ as $|\mathbf{k}|\rightarrow0$. This stands in stark contrast with the established paradigm of critical phenomena, wherein the longest timescale governs the evolution. This apparent paradox resolves upon recognizing that, although the dissipation is negligibly small within the coherent timescale $|\mathbf{k}|^{-z}$, its cumulative action can no longer be ignored on the extended scale $\lambda^{-1}|\mathbf{k}|^{-sz}$. Thus the late-time dynamics inevitably restores the rule of the slowest degree of freedom, notwithstanding the naive verdict of irrelevance delivered by power counting. We refer to this restoration of canonical late-time governance as dissipation-regression scaling (DRS). Accordingly, Eq.~(\ref{sgtb}) is superseded by
\begin{equation}
	{\mathfrak D}(t,g,T,\lambda,L)= b^{-d}{\mathfrak D}(t{\lambda} b^{-sz}, gb^{1/\nu}, Tb^{z}, L^{-1}b^{-1}),\label{sgtbr}
\end{equation}
where} $\lambda$ enters merely as a {dimensionless} time unit, in analogy to the kinetic coefficient in the Langevin equation for critical dynamics~\cite{Hohenberg,Folk}.	{Note that DRS lies in the coherent regime where the dynamic critical exponent $z$ remains unchanged. We emphasize that this DRS represents a conceptually new mechanism, fundamentally differing from the usual crossovers between different regimes as exemplified by Eq.~(\ref{sgtla}) and from more intricate scenarios involving multiple competing scales~\cite{Cao}.} 

{For $s<2$, the dissipation timescale exceeds its coherent counterpart, indicating that dissipation is dominant. This gives rise to the dissipative universality class~\cite{Werner,Sper,Sper12,Oshiyama}. The dominant $\omega^s$ implies $\omega\sim\lambda^{-1/s}\mathbf{k}^z$, so that $z$ takes the value $2/s$ and Eq.~(\ref{sgtb}) changes into
	\begin{equation}
		{\mathfrak D}(t,g,T,\lambda,L)= b^{-d}{\mathfrak D}(t\lambda^{-1/s} b^{-z}, gb^{1/\nu}, Tb^{z}, L^{-1}b^{-1}).\label{sgtbrs}
	\end{equation}
similar to the classical critical phenomena with memory~\cite{Zeng,Zeng1}, where $\lambda$ serves once again as a time unit. Indeed, while an Ohmic bath yields the customary linear dissipation---i.e., local first-order time derivative---in the real-time domain~\cite{Verstr,Quan,Sieberer}, a generic sub-Ohmic bath introduces a fractional temporal derivative~\cite{Samko,Metzler,Zaslavsky,West,Tarasovb} equivalent to a non-local long-range temporal interaction~\cite{Verstr}, in exact parallel with the classical case~\cite{Zeng,Zeng1}. Note that Eq.~(\ref{sgtbrs}) is essentially Eq.~(\ref{sgtb}), except for different values of critical exponents~\cite{Werner,Sper12} and the absence of $\lambda$ as an irrelevant variable. Note also that, in contrast to what happens in the coherent regime, although the coherent timescale exceeds its dissipation counterpart, dissipation kills the evolution before coherence can manifest.} 

{These mean-field results continue to hold upon inclusion of fluctuations. Standard dimensional analysis of Eq.~(\ref{sphi}) assigns a scaling dimension $4-(d+z)$ to $u$. This implies an upper critical dimension of $d_c=4-z$ below which fluctuations become relevant, in agreement with the quantum classical correspondence~\cite{Hertz,sachdev}. Owing to the fluctuation-dissipation relation exhibited in Eq.~(\ref{gram1}), no systematic spatiotemporal dimension shift is needed~\cite{Zeng,Zeng1}. An $\varepsilon=d_c-d$ perturbative RG analysis can then confirm that these two regimes are governed by two fixed points. A consequence is that both $z$ in the dissipative regime and the locus of the crossover between the two regimes, $s=s_x$, receive their leading corrections at order $\varepsilon^2$, as in the corresponding classical problems~\cite{Zeng1,Bonart,Pankov}, although the theory here is different in that it is inherently quantum and real-time. However, Lorentz-invariance guarantees $z=1$ throughout the coherent regime. A full RG treatment goes beyond the scope of this work and is deferred to a future publication. Suffice it to note here that Monte Carlo simulations of the discretized open quantum Ising chain show $z$ increasing monotonically from $1$
toward approximately $2$ as $s$ decreases from $2$ to $1$~\cite{Werner,Oshiyama,Sper12,Sper}. These results rule out $s_x=1$ since $z$ would then equal $1$ rather than $2$. Note that in mean field, $s_x=2$, where $z=1$ in both regimes.}
	
{An apparent tension emerges between the field-theoretic construction, Eqs.~(\ref{qft})--(\ref{gram1}), and the Lindblad equation, Eq.~(\ref{drho}), regarding the location of the marginal bath-exponent $s$: the former predicts $s_x=2$ while the latter gives $s_x^{\rm Lind}=1$, as signaled by the distinct $\lambda$ RG eigenvalues obtained within each scheme, despite both purporting to capture criticality of the same underlying model. The two frameworks cannot be trivially reconciled: numerical results mentioned demonstrates $s_x>1$ when fluctuations are taken into account. On the one hand, $s_x>1$ indicates that even super-Ohmic baths of $1<s<s_x$ can alter the critical physics; on the other hand, the unrenormalized boundary at $s_x^{\rm Lind}=1$ remains broadly compatible with the expectation that only sub-Ohmic baths affect universal properties. Although different forms of coupling to the bath may change $s_x$~\cite{Sper}, the main reason lies in the dissimilar low-energy spectra assumed by each side to realize $z=1$. Specifically, whereas the frequency and wave number in the quantum field theory of Eq.~(\ref{qft}) enter quadratically, applying the same framework naively to the diagonalized Hamiltonian generates a propagator closely akin to $\mathcal{G}_R^{-1}$, except for a linear coherent pole $\omega-\epsilon_{\mathbf{k}}$ with $\epsilon_{\mathbf{k}}\propto|\mathbf{k}|$ near criticality. This replaces the boundary $s_x=2$ by $s_x^{\rm Lind}=1$. Fluctuation corrections could plausibly renormalize this value downward as the case of $s_x$ does. This implies that $s=1$ itself sits at least on the boundary of the coherent regime governed by DRS.}

{The mismatch between the boundary location $s_x^{\rm Lind}$ and its discretized counterpart $s_x$---together with the distinctly different $z$ values attained at least for $s_x^{\rm Lind}<s<s_x$---signals an intrinsic disparity in the dynamical content of the two formulations applied to the same Hamiltonian. Such a multiplicity has a well-established precedent in classical critical dynamics, where Model A and Model B belong to separate dynamic universality classes despite sharing the same static fixed point~\cite{Hohenberg}. By analogy, the two constructions examined here should likewise be recognized as genuine representatives of two distinct dynamic universality classes, even though neither additional conservation law nor auxiliary fields are invoked. This suggests that the dispersive encoding of the quasiparticle spectrum warrants recognition as an additional factor in current classifications of dynamic critical phenomena. As will become evident below, DRS likewise modifies the apparent dynamical exponent, thereby altering, in effect, the dynamic universality class. Given the sensitivity just highlighted, this change of universality class should no longer be regarded as unusual.}

{We note that, mirroring the discretized case, a transition at some $s_x^{\rm Lind}<1$ seems inherent to the diagonalized models, although its dissipative wing still awaits verification. Fortunately, our reasoning never depends on that possibility: as we will demonstrate below, in all instances with $s\geq1$, numerical outcomes place the diagonalized model safely within the coherent regime described by DRS. Together with earlier numerical results locating the discretized variant in the dissipative regime across $1\leq s<2$~\cite{Werner,Oshiyama,Sper12,Sper}, it is unequivocal that the two representations belong to different universality classes at least over that common span.}

{For the special case $s=0$, it might lead to a diverging $z$ regardless of whether $z=1/s$ or $z=2/s$. However, as mentioned and seen in Eq.~(\ref{sgtb}), at $s=0$, $\lambda$ has the dimension of $\omega$ or $\omega^2$ in the two cases and thus is as dominant as the latter. Moreover, $\omega^0=1$ and the time dependence of the dissipation term is lost. Consequently, $z$ can only arise from the coherent evolution. On these grounds, $s=0$ necessarily satisfies DRS.}

{Note that the separation of coherent and dissipative regime does not depend on the magnitude of $\lambda$. Indeed, from Eq.~(\ref{gram1}), in the coherent regime, $\omega^2\gg\omega^s$ as $\omega\rightarrow0$ for $s>2$, provided that $\lambda$ stays finite. Only in a finite-sized system in which a nonzero gap $\omega_g$ survives at criticality can a sufficiently large $\lambda>\omega_g^{2-s}$ overturn the inequality and pull the dynamics into the dissipative sector. Conversely, for $s<2$, a weak-enough $\lambda<\omega_g^{2-s}$ can switch from the dissipative regime to the coherent one and initiate DRS.}

{We now deduce concrete} predictions from {DRS encapsulated in}, Eq.~(\ref{sgtbr}){, and confront them with recent numerical data. A first immediate result follows by fixing $b=(\lambda t)^{1/sz}$ to find}
\begin{equation}
	{\mathfrak D}(t,g,T,\lambda)= ({\lambda} t)^{-d/sz}f_t(g({\lambda}t)^{1/s\nu z}, T({\lambda}t)^{1/s}),\label{sgt}
\end{equation}
where $f_t$ is yet another scaling function and{, for brevity, the system size has been suppressed (restored straightforwardly when needed)}. Equation~(\ref{sgt}) {predicts a universal algebraic decay of the form} $({\lambda}t)^{-d/sz}$ exactly at the quantum critical point $g=T=0$, in {full} agreement with the numerical results {for the Kitaev chain at $s=1$, $2$, and $3$, along both the critical line ($z=1$) and the multicritical point ($z=2$), see Fig.~10 in ~\cite{Ding}}. {By contrast,} in the standard setup of Eq.~(\ref{sgtb}), setting $b=t^{1/z}$ recovers the purely coherent evolution ${\mathfrak D}\sim t^{-d/z}${, which continues to hold at the dissipative fixed point in Eq.~(\ref{sgtbrs}) albeit with a different $z=1/s$ in mean field. Remarkably, neither of these two situations permits the particular combined exponent $sz$ to emerge anywhere. This observation offers striking confirmation of the proposed theory of DRS.} Away from the critical point, the validity of Eq.~(\ref{sgt}) requires $g({\lambda}t)^{1/s\nu z}\ll1$ and $T({\lambda}t)^{1/s}\ll1$, or equivalently, $t\ll{\lambda^{-1}}|g|^{-s\nu z}\sim\zeta_{\rm corr}$ and $t\ll{\lambda^{-1}} T^{-s}$. These two expressions are {transparent} because they state that the time must be much smaller than the equilibrium correlation time $\zeta_{\rm corr}$ and the imaginary time $T^{-s}$; otherwise the system would have {practically} equilibrated. {At first sight it may seem surprising that both characteristic timescales acquire an additional factor of $s$ in their defining exponents. Yet this is an inevitable structural consequence of the new timescale introduced in Eq.~(\ref{sgtbr}): once the timescale itself carries a dimension proportional to $s$, every composite timescale inherits the same factor. Heuristic versions of precisely these timescales have been invoked to phenomenologically derive the specific Kibble–Zurek scaling~\cite{Bacsi,Kuo,Ding}. Within the systematic framework developed here, however, they emerge on firmer ground, governing all remaining scaling relations in the problem, rather than assumed by hand.}

{Note that Eq.~(\ref{sgt}) is ill-defined at $s=0$. In fact, in this case, ${\mathfrak D}$ decays exponentially~\cite{Ding}, only residual scaling remains~\cite{Yinchen}.}

We proceed to {drive the system by sweeping either $g$ according to Eq.~(\ref{grt}) or $T$ according to}
\begin{equation}
	T=\pm R_Tt^m\label{Trt}
\end{equation}
with constants $R_T>0$ and $m$, where $\pm$ applies to heating or cooling. {Again assuming the transformed quantities follow the same equations as Eqs.~(\ref{grt}) and~(\ref{Trt}), we find $g'=gb^{1/\nu}=Rt^nb^{1/\nu}=R't^n\lambda^{n}b^{-nsz}$ from Eqs.~(\ref{grt}) and~(\ref{sgtbr}) and $T'=Tb^{z}=R_Tt^mb^{z}=R_T't^m\lambda^mb^{-msz}$ from Eqs.~(\ref{Trt})} and~(\ref{sgtbr}), resulting in
\begin{equation}
	r_g=nsz+1/\nu,\qquad r_T=(1+ms)z.\label{rgnsz}
\end{equation}
{By contrast, the standard Eq.~(\ref{sgtb}) and its dissipative counterpart Eq.~(\ref{sgtbrs}) produce instead $r_T=(1+m)z$, which exhibits a fundamentally different dependence on $s$ for $z=1/s$. The same difference also manifests itself in Eq.~(\ref{rznu}) for sweeping $g$.} 
Again replacing $t$ with $R$ or $R_T$ transforms Eq.~(\ref{sgtbr}) into
\begin{eqnarray}
	{\mathfrak D}(R,g,T,\lambda)= b^{-d}{\mathfrak D}(\lambda^{-n}Rb^{r_g}, gb^{1/\nu}, Tb^{z}),~\label{sgtbrr}\\
	{\mathfrak D}(R,g,T,\lambda)= b^{-d}{\mathfrak D}(\lambda^{-m}R_Tb^{r_T}, gb^{1/\nu}, Tb^{z}).\label{sgtbrT}
\end{eqnarray} 
Setting $b=(\lambda^{-n}R)^{-1/r_g}$ or $b=(\lambda^{-m}R_T)^{-1/r_T}$ then yields the FTS forms for the two driving protocols
\begin{eqnarray}
	{\mathfrak D} =(\lambda^{-n}R)^{d/r_g}f_g(g\lambda^{-n}R^{-1/r_g\nu}, T\lambda^{-n}R^{-z/r_g}),\qquad~\label{sgrr}\\
	{\mathfrak D} =(\lambda^{-m}R_T)^{d/r_T}f_T(g\lambda^{-m}R_{{T}}^{-1/r_T\nu}, T\lambda^{-m}R_{{T}}^{-z/r_T}),\quad~~\label{sgTr}	
\end{eqnarray} 
where $f_g$ and $f_T$ are universal scaling functions. Consequently, using Eq.~(\ref{rgnsz}), we arrive at the Kibble-Zurek scaling exponents
\begin{equation}
	\sigma_g=d\nu/(1+nsz\nu),\quad \sigma_T=d/(1+ms)z,\label{sg}
\end{equation}
both of which {again exhibit a unique $s$ dependence. Remarkably, these predictions accurately reproduce published numerical results, as we now demonstrate.} 

{Independent support for these scaling forms and exponents comes from several sources. First, inverting the first argument on the right-hand side of Eq.~(\ref{sgtbrT}) reproduces precisely the very same} homogeneous {scaling} function {presented as Eq.~(8) of}  Ref.~\cite{King}{, except for an extra parameter representing the initial temperature $T_i$. In the present work, we deliberately choose initial configurations far enough away from criticality to make their memory fully negligible; wherever required, they can be systematically handled within FTS~\cite{Feng}. Eq.~(8) of Ref.~\cite{King} rested on an} exact rate-equation {for a specifically Kitaev-wire setting ($s=1$ and $m=1$), accompanied by successful numerical checks. Here we recover it on substantially wider grounds, free of restrictions to any particular model or Markovian assumption.
Crucially, this precise functional form cannot be obtained from the conventional form Eq.~(\ref{sgtb}), from the dissipation-dominated form (\ref{sgtla}), nor from any other ansatz built upon an alternative control variable. 
Second, $\sigma_T$ in Eq.~(\ref{sg}) maps exactly onto the heuristically motivated result of} Ref.~\cite{Bacsi}, {upon identifying our $R$ with their $\tau^m$ and changing $m$ to their $p$. Third, more strongly, both $\sigma_g$ and $\sigma_T$ in Eq.~(\ref{sg}) are identical to those phenomenologically derived and numerically validated in} Ref.~\cite{Kriel}. {Fourth, in addition to the above analytic and numerical evidence, direct numerical solutions of the open quantum Ising chain with $z=1$ yield $\sigma_T=1$, $1/2$, and $1/3$, for linear temperature ramps ($m=1$) at $s=0$, $1$, and $2$, respectively~\cite{Bacsi}, in precise agreement with Eq.~(\ref{sg}). Fifth, the precise combination of $\lambda$ and $R$ in Eq.~(\ref{sgTr})  match exactly the numerical outcomes~\cite{Bacsi}. We note that $\lambda$ serves only as a time unit, as is the case of $s=1$ in Eq.~(\ref{sgtb}). Hence, the apparent divergence encountered as $\lambda\rightarrow0$ is not indicative of a physical instability or a failure of the theory. It arises simply from the progressive shrinking of the time unit employed in the definition of the dimensionless variables.}

{Sixth, more remarkably, the same dictionary translates essentially unaltered into a wholly separate physical environment, namely quantum non-demolition measurement-induced decoherence. There, the approximate Markovian master equation takes the form of Eq.~(\ref{drho}) with ${\cal L}$ replaced by ${\cal H}$~\cite{Kuo}. Since each ${\cal H}$ carries the energy dimension one, the composite dissipator behaves as $\omega^2$, thereby locating this superficially disconnected realization exactly at $s=2$. For a linear ramp of $g$, the present framework directly produces the frozen correlation length evaluated at $\hat{g}$ as $\hat{\xi}\sim(\lambda^{-1}R)^{-1/r_g}\sim (\lambda/R)^{\nu/(1+2z\nu)}$, while the freeze-out time is $\hat{t}\sim\lambda^{-1} (\lambda^{-1} R)^{-sz/r_g}=(\lambda^{-1}R^{-2z\nu})^{1/(1+2z\nu)}$, obtained by substituting the scaling factor $b$ into the timescale of Eq.~(\ref{sgtbr}). Upon identifying their quench rate $1/\tau$ with our $R$, both expressions coincide exactly with the central results of Ref.~\cite{Kuo}. Notably, $\hat{t}\sim\lambda^{-1}\hat{\xi}^{sz}$ inherits once again the additional power of $s$ that distinguishes the present framework. A scaling collapse in Ref.~\cite{Kuo} further substantiates this agreement, tying the phenomenology back to the unifying DRS picture developed here.}

{We have seen that different real-time dynamics can alter the universality class for $s<2$ once the open systems enter the dissipative regime. Moreover, an extra $s$ factor appears in the} time-related quantities such as the correlation time and $T$, as exemplified by the consequences derived from Eq.~(\ref{sgt}) {for $s>2$}. From the perspective of universality class, this change in exponents signifies {again} a change of dynamic universality class with varying $s$, except for $s=1$. Nevertheless, from the standpoint of determining the universality class of the original QPT, it can be straightforwardly recovered from the relations among the exponents once $s$ is known.

In summary, we have {formulated dissipation-regression scaling (DRS), which systematically reproduces} the observed unusual Kibble-Zurek scaling {as well as relaxation behavior}, capturing not only the leading scaling exponent itself, but also other {subleading} contributions. {Our approach builds upon unraveling two subtleties in the quantum criticality of open systems coupled to their finite-temperature environment. Through a nonequilibrium Schwinger-Keldysh field theory for a low-energy canonical bosonic model, we have shown that the bath spectral exponent $s$ separates the problem into a coherent regime with unitary dynamics for $s>2$ and a dissipative regime featuring a fresh set of critical exponents for $s<2$, consistent with numerical simulations of the discretized open quantum Ising model. 
The key subtlety is that the longest timescale within both regimes belongs not to the RG-relevant operator but to the nominally irrelevant one. Focusing on the coherent regime, the dissipation coupling, endowed with its distinctive temporal scaling dimension, is the irrelevant variable that determines the longest timescale and therefore necessarily dominates at late times, yielding the DRS. At least within open quantum criticality, this illustrates that scaling dimensions alone are insufficient for determining an operator’s impact on late-time dynamics.}

{Equally striking is that applying the same field-theoretic framework to the diagonalized variant of the open quantum Ising chain (described by a Lindblad master equation) yields a universality class inequivalent to its directly discretized counterpart for $s<2$. The origin of this divergence lies in how $z=1$ is realized: either through the analytic structure of the low-energy spectrum or through the differential order of the time evolution generator. Existing numerical results have convincingly established the reality of this disparity, indicating that the dispersive encoding of the quasiparticle spectrum merits inclusion as an additional principle in the taxonomy of dynamic critical phenomena.}
	
{DRS applies to the coherent regime for $s\geq2$ or $s\geq1$ in the discretized and diagonalized models, respectively. Its fingerprint is a multiplicative rescaling of the dynamic exponent by $s$. Consequently, a new dynamic universality class emerges irrespective of the coupling strength, although} the critical exponents of the original QPT can be {readily reconstructed. These results illustrate the acute sensitivity of universality classes to the real-time dynamics and the richness of criticality in open quantum systems.} 

We {have shown that the perturbation-expansion result} $\sigma_{\rm inc}$, Eq.~(\ref{sinc}), {is valid for the two} special cases $s=1$ (Ohmic) and $s=0$ (with $n\nu=1$). Significantly, it follows from expanding the correlation length rather than the excitation density itself, exposing a structural limitation inherent to using the latter as the starting point for perturbative treatments. Moreover, for a nonlinear driving, the same perturbative treatment of the Ohmic case fails to reproduce the correct $\lambda$-dependence, indicating that the coupling $\lambda$ cannot, in general, be treated as a perturbation {even if it is} weak.

Several intriguing questions arise for future study. We have seen that the DRS modifies the critical exponents of some quantities. However, in the critical phenomena with memory~\cite{Zeng,Zeng1}, the long-range temporal interaction of the system radically alters its critical behavior, even at the mean-field level, by introducing a series of brand new phenomena such as violation of scaling laws, spatiotemporal dimension shifts, discontinuous crossover from short-range to long-range behavior, anomalous system-size dimensions, and unusual behaviors above the upper critical dimension, in sharp contrast to the case of long-range spatial interaction~\cite{Fisher,Sak}. It is then desirable to know whether such behaviors can emerge here in the open quantum systems. {Another} question is related to new exponents in the driven open quantum systems we consider. Although we have new critical exponents arising from the coupling at $s\geq1$, they are all determined by the known equilibrium and dynamic critical exponents, as well as $s$. The question here is whether fundamentally new exponents different from the foregoing ones can emerge in such driven open quantum systems. Some new exponents have been found by breaking self-similarity symmetry determined by $L^{-1}R^{-1/r}$ as appears in Eq.~(\ref{sgtr})~\cite{Yuan,Yuan1,Yuan2}: If $L^{-1}R^{-1/r}$ is fixed such that systems of different sizes contain the same number of driving lengths $R^{-1/r}$, i.e., self-similarity symmetry holds, then no new exponents are needed; whereas if it is broken, new exponents must be invoked. Note that these new exponents stem from critical phenomena rather than phase ordering~\cite{Biroli,Jeong}. {Yet other desirable investigations include numerical explorations of the discretized open quantum Ising models for $s\geq2$, and separately, of their diagonalized counterparts for $s<1$, and the relevance of the theory to broader classes of open quantum criticality such as anti-Kibble-Zurek scaling whereby slower driving leads, counter-intuitively, to more defects ~\cite{Griffin,Dutta,Gao,Puebla,Ai,Iwamura,Kou,Du}.}

\begin{acknowledgments}
This work was supported by National Natural Science Foundation of China (Grant No. 12175316).
\end{acknowledgments}

\end{document}